# Classification of Normal/Abnormal Heart Sound Recordings based on Multi-Domain Features and Back Propagation Neural Network

Hong Tang[1], Huaming Chen [1], Ting Li[2], Mingjun Zhong[1]

[1]Dalian University of Technology, Dalian, China
[2] Dalian Minzu University, Dalian, China

## Abstract

*This paper aims to classify a single PCG recording as normal or abnormal for computer-aided diagnosis. The proposed framework for this challenge has four steps: preprocessing, feature extraction, training and validation. In the preprocessing step, a recording is segmented into four states, i.e., the first heart sound, systolic interval, the second heart sound, and diastolic interval by the Springer Segmentation algorithm. In the feature extraction step, the authors extract 324 features from multi-domains to perform classification. A back propagation neural network is used as predication model. The optimal threshold for distinguishing normal and abnormal is determined by the statistics of model output for both normal and abnormal. The performance of the proposed predictor tested by the six training sets is sensitivity 0.812 and specificity 0.860 (overall accuracy is 0.836). However, the performance reduces to sensitivity 0.807 and specificity 0.829 (overall accuracy is 0.818) for the hidden test set.*

## 1. Introduction

Heart sounds provide important initial clues in cardiovascular disease evaluation. It is necessary to develop automatic algorithms for computer-aid diagnosis. The previous studies used frequency spectrum [1], wavelet coefficients [2] to classify heart sound recordings based on self-organization map, grow and learn network, linear vector quantization. However, the recordings in these studies were limited by the recording number, length, signal frequency range, or the environments of data collection. To address the problem, the physioNet/Cinc Challenge 2016 provides a large collection of heart sound recordings for this purpose [3]. The authors extract multi-domain features to perform classification.

## 2. Methods

### 2.1. Descriptions of the heart sound databases

The opened training set consists of six databases (A through F) containing a total of 3153 PCG recordings, collected at either a clinical or nonclinical environment, from both healthy subjects and pathological patients. Each recording has been manually labeled as normal (-1) and abnormal (1). The training and test sets have each been divided so that they are two sets of mutually exclusive population. The details of the databases and challenge can be found in [4].

### 2.2. Preprocessing

A heart sound signal is filtered with pass band [20 120] Hz to remain the main part of heart sounds. Then it is inversely filtered to avoid group delay. Springer's segmentation algorithm [5] is applied to the filtered signal for segmenting. The heart sound signal sequence is labeled by four states (S1, systole, S2 and diastole) for each beat. The feature extraction is therefore carried on based on the state labels.

### 2.3. Extracted features

In this study, 324 features are extracted from each recording from multi-domains. They are listed in detail in the follows.

**22 Features in domain of heart sound intervals**

Liu et al [4] proposed 20 features in heart sound intervals.
1. m_RR: mean value of RR intervals
2. sd_RR: standard deviation (SD) of RR intervals
3. m_IntS1: mean value of S1 intervals
4. sd_IntS1: SD of S1 intervals
5. m_IntS2: mean value of S2 intervals
6. sd_IntS2: SD of S2 intervals
7. m_IntSys: mean of systolic intervals
8. sd_IntSys: SD of systolic intervals

9. m_IntDia: mean of diastolic intervals
10. sd_IntDia: SD of diastolic intervals
11. m_Ratio_SysRR: mean of the ratio of systolic interval to RR of each heart beat
12. sd_Ratio_SysRR: SD of the ratio of systolic interval to RR of each heart beat
13. m_Ratio_DiaRR: mean of ratio of diastolic interval to RR of each heart beat
14. sd_Ratio_DiaRR: SD of ratio of diastolic interval to RR of each heart beat
15. m_Ratio_SysDia: mean of the ratio of systolic to diastolic interval of each heart beat
16. sd_Ratio_SysDia: SD of the ratio of systolic to diastolic interval of each heart beat
17. m_Amp_SysS1: mean of the ratio of the mean absolute amplitude during systole to that during the S1 period in each heart beat
18. sd_Amp_SysS1: SD of the ratio of the mean absolute amplitude during systole to that during the S1 period in each heart beat
19. m_Amp_DiaS2: mean of the ratio of the mean absolute amplitude during diastole to that during the S2 period in each heart beat
20. sd_Amp_DiaS2: SD of the ratio of the mean absolute amplitude during diastole to that during the S2 period in each heart beat
  The authors add another two features:
21. m_Ratio_IntS1S2: mean value of the ratio of S1 interval to S2 interval
22. sd_Ratio_IntS1S2: SD of the ratio of the mean.

**10 Features in energy domain**
1. Ratio_energy_HSTotal: Ratio of the sum of heart sounds' energy to the total energy of a recording
2. Ratio_magnitude_HSTotal: Ratio of sum of heart sounds' absolute magnitude to the sum of absolute magnitude of a recording
3. Ratio_energy_HSRemain: Ratio of the sum of heart sounds' energy to the energy of the remain parts
4. Ratio_magnitude_HSRemain: Ratio of the sum of heart sounds' absolute magnitude to the sum of the absolute remain parts
5. m_Ratio_energy_SysCycle: mean value of the ratio of energy in systolic interval to cardiac cycle energy of each heart beat
6. sd_energy_SysCycle: SD of the ratio of the mean
7. m_Ratio_energy_DiaCycle: mean value of the ratio of energy in diastolic interval to cardiac cycle energy of each heart beat
8. sd_energy_DiaCycle: SD of the ratio the mean
9. m_Ratio_HSCycle: mean value of the Heart sounds energy to cardiac cycle energy of each heart beat
10. sd_energy_HSCycle: SD of the ratio of the mean

**82 Features in frequency spectrum**
1-12. m_Fre_Spec_S1_f: mean of frequency spectral values of S1 at frequency $f$. The frequency $f$ is considered from 10 Hz to 120 Hz with interval of 10 Hz
13-24. m_Fre_Spec_S2_f: mean of spectral values of S2 at frequency $f$. The frequency $f$ is considered from 10 Hz to 120 Hz with interval of 10 Hz.
25-53. m_Fre_Spec_Sys_f: mean of spectral values of systole signal at $f$ of each heart beat where the frequency is from 10 Hz to 290 Hz with interval of 10 Hz
54-82. m_Fre_Spec_Dia_f: mean of spectral values of diastole of each heart beat from 10 Hz to 290 Hz with interval of 10 Hz

**2 Features in heart rate sequence**
1. m_HR: Mean of heart cycle period. The method to calculate m_HR is different from that to calculate m_RR. The heart rate is based on analysis of the autocorrelation function and the positions of the peaks therein.
2. sd_HR: SD of the mean

**57 Features in frequency spectrum of heart rate sequence**
1-19. spec_HR_seq_f: Spectral values of heart rate sequence from 0.05 Hz to 1 Hz with interval 0.05 Hz. The heart rate sequence is non-uniformly sampled in time domain due to heart rate variability. So, nonlinear interpolation by 'cubic' method is used in this study.
20-38. spec_Sys_seq_f: Spectral values of systole sequence from 0.05 Hz to 1 Hz with interval 0.05 Hz.
39-57. spec_Dia_seq_f: Spectral values of diastole sequence from 0.05 Hz to 1 Hz with interval 0.05 Hz.

**8 Features in Kurtosis**
1. m_S1_kurtotsis: mean values of the kurtosis of S1
2. sd_S1_kurtosis: SD of the mean
3. m_S2_kurtotsis: mean values of the kurtosis of S2
4. sd_S2_kurtosis: SD of the mean
5. m_Sys_kurtotsis: mean values of the kurtosis of systole signal
6. sd_Sys_kurtosis: SD of the mean
7. m_Dia_kurtotsis: mean values of the kurtosis of diastole signal
8. sd_Dia_kurtosis: SD of the mean
  The kurtosis of a signal $x(t)$ is calculated by the formula

$$K_x = \frac{E[x^4(t)]}{(E[x^2(t)])^2}, \qquad (1)$$

where $E(.)$ is an expectation operator.

**4 Features in cyclostationarity**
1. m_cyclostationarity_1: mean value of the degree of cyclostationarity. The definition of "degree of cyclostationarity" can be found in [6]. This feature indicates the degree of signal repetition. It will be infinite if the events occurred in heart beating were exact periodic.

However, it will be a small number if the events are random alike. Let's assume $\gamma(\alpha)$ is the cycle frequency spectral density of a heart sound signal at cycle frequency $\alpha$. This feature is defined as

$$d(\eta) = \gamma(\eta)/\int_0^\beta \gamma(\alpha)d\alpha \qquad (2)$$

where $\beta$ is the maximum cycle frequency considered and $\eta$ is the basic cycle frequency indicated by the first peak location of $\gamma(\alpha)$. A heat sound signal is equally divided into subsequence. The feature can be estimated for each subsequence, then the mean and standard deviation can be obtained.

2. sd_cyclostationarity_1: SD of the mean
3. m_cyclostationarity_2: The definition of this indicator is the sharpness of the peak of cycle frequency spectral density. It is

$$\text{peak\_sharpness} = \max(\gamma(\alpha))/\text{median}(\gamma(\alpha)). \qquad (3)$$

The operators, max(.) and median(.) are the maximum magnitude and median of the cycle frequency spectral density. It is obviously that the sharper of the peak is, the greater the feature is. Similarly, the feature can be calculated for each sub-sequence of the heat sound signal and then get the mean and SD.
4. sd_cyclostationarity_2: SD of the mean

**82 Features in power spectral density**
1-12. m_Pow_Spec_S1_f: mean of frequency spectral values of S1 at frequency $f$ of each heartbeat. The frequency $f$ is considered from 10 Hz to 120 Hz with interval of 10 Hz
13-24. m_Pow_Spec_S2_f: mean of spectral values of S2 at frequency $f$ of each heartbeat. The frequency $f$ is considered from 10 Hz to 120 Hz with interval of 10 Hz.
25-53. m_Pow_Spec_Sys_f: mean of spectral values of systole signal at f of each heart beat where the frequency is from 10 Hz to 290 Hz with interval of 10 Hz
54-82. m_Pow_Spec_Dia_f: mean of spectral values of diastole of each heart beat from 10 Hz to 290 Hz with interval of 10 Hz

**57 Features in power spectral density of heart rate sequence**
1-19. Pow_spec_HR_seq_f: power spectral values of heart rate sequence from 0.05 Hz to 1 Hz with interval 0.05 Hz
20-38. Pow_spec_Sys_seq_f: power spectral values of systole interval sequence from 0.05 Hz to 1 Hz with interval 0.05 Hz
39-57. Pow_spec_Dia_seq_f: power spectral values of diastole interval sequence from 0.05 Hz to 1 Hz with interval 0.05 Hz

### 2.4. Prediction model

A back propagation artificial neural network is applied to approximate the link between the features and blood pressure, whose structure with two hidden layers is shown in Fig. 1. A back propagation neural network is a feed-forward network with its weights adjusted through the method of back propagation learning algorithm and it can achieve arbitrary nonlinear mapping from input to output, generally having a good performance. Each net node in the network is a neuron whose function is to calculate the inner product of the input vector and weight vector by a nonlinear transfer function to get a scalar result. The "logsig" function is applied to the hidden layer 1. The "purelin" function is applied to the hidden layer 2 and the output layer. The number of the neurons for the hidden layer 1 and hidden layer 2 are empirically chosen to be 10 and 5, respectively. The training algorithm applied is the "trainlm".

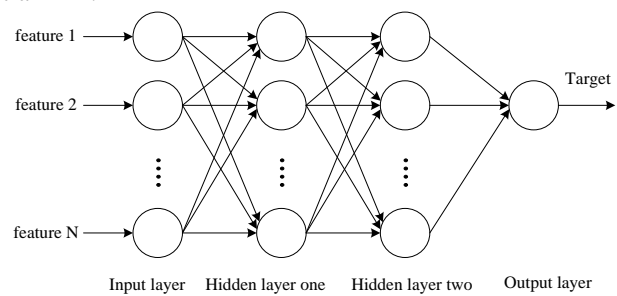

Fig. 1. Structure of the neural network

### 3. Results

### 3.1. Threshold to classify normal and abnormal

The prediction model outputs a real number. According the link trained by the input-output pairs, the output of the prediction model will approach -1 if a recording is normal and approach 1 if a recording is abnormal. To determine an optimal threshold for classifying, it is necessary to investigate the statistic of outputs for both normal and abnormal. The all inputs are selected to train the model, histogram analysis is performed for the outputs of both normal and abnormal recordings, see the arrow in Fig. 2. It can be found that the cross point of the two curves is at about -0.4. However, the position of the cross point changes if the percent number of input varies. To investigate this change, firstly, 50% of the inputs were randomly selected to feed into the model, and then 60% of inputs were to put into the model, and repeat the operation with increment 10% each time. The position of the cross point with respect to the percent inputs is listed in Table 1. It is found that the position of the cross point varied little with the percent number. So, the threshold is finally set to the average, -0.52. Those recordings whose model output greater than the threshold are classified as abnormal. To the contrary, those recordings whose output less or equal to the threshold are classified as normal.

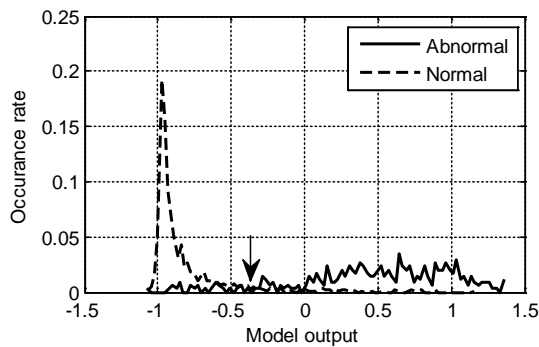

Fig. 2. Histogram of the model output for normal and abnormal

Table 1. Position of the cross point with respect to percent number of inputs

| Percent number | Position of cross point |
|---|---|
| 50% | -0.5 |
| 60% | -0.6 |
| 70% | -0.6 |
| 80% | -0.5 |
| 90% | -0.5 |
| 100% | -0.4 |
| Average | --0.52 |

### 3.2. Performance of classification

The training set is exclusively divided into two parts. The first part is used to train the model and the other part is used to test the model. The percent to train is from 30% to 90% of the training set, and the performances were listed in table 2. It is found that the prediction performance of the model is lightly increasing with the percent to train.

Table 2 Performance of the model tested in various situations.

| Percent to train | Percent to test | Sensitivity | specificity | Overall |
|---|---|---|---|---|
| 30% | 70% | 0.712 | 0.854 | 0.783 |
| 40% | 60% | 0.817 | 0.848 | 0.832 |
| 50% | 50% | 0.823 | 0.810 | 0.816 |
| 60% | 40% | 0.874 | 0.753 | 0.813 |
| 70% | 30% | 0.860 | 0.780 | 0.820 |
| 80% | 20% | 0.830 | 0.836 | 0.833 |
| 90% | 10% | 0.812 | 0.860 | 0.836 |

Best prediction performance out of ten attempts for the hidden set given by the organization committee is: sensitivity 0.807, specificity 0.829, overall score 0.818.

### 4. Conclusions

This paper tries to classify heart sound signal recordings into normal or abnormal based on multi-domain features. 324 features were extracted to train a back propagation neural network for the prediction. The best performance of the model based on training set is: sensitivity 0.812, specificity 0.860, overall score 0.836. However, the performance tested by the hidden set is: sensitivity 0.807, specificity 0.829, overall score 0.818. The authors believe that the performance would be improved if a prediction model based on deep learning is used.

### Acknowledgements

This work was partially funded by the National Natural Science Foundation of China under Grant 61471081 and Fundamental Research Funds for the Central Universities under Grant No. DUT15QY60, DUT16QY13.

Address for correspondence.
Hong Tang
Department of Biomedical Engineering, Room A1213, School of Electronic Information and Electrical Engineering, Dalian University of Technology, Linggong Road #2, Dalian city, 116024, China
tanghong@dlut.edu.cn